\documentclass[a4paper,backend=bibtex]{jacow}
\graphicspath{{figures/}}
\usepackage[USenglish]{babel}			 
\usepackage[final]{pdfpages}
\usepackage{authblk}

\begin{document}

\title{Precision beam energy measurement by undulator radiation at
  MAMI\thanks{Work supported partly by Deutsche Forschungsgemeinschaft
    (DFG) in the framework of GRK 2128 and PO-256/7-1, by DAAD PPP
    57345295/JSPS, by JSPS KAKENHI No.~JP17H01121 and by Tohoku
    University (GP-PU).}}

\author[1]{P.~Klag\thanks{klagp@uni-mainz.de}}
\author[1]{P.~Achenbach}
\author[1]{M.~Biroth}
\author[2]{T.~Gogami}
\author[1]{P.~Herrmann}
\author[2]{M.~Kaneta}
\author[2]{Y.~Konishi}
\author[1]{W.~Lauth}
\author[2]{S.~Nagao}
\author[2]{S.~N.~Nakamura}
\author[1]{J.~Pochodzalla}
\author[1]{J.~Roser}
\author[2]{Y.~Toyama}

\affil[1]{Institut f\"ur Kernphysik, Johannes Gutenberg-Universit\"at,
  55099 Mainz, Germany}

\affil[2]{Department of Physics, Tohoku University, Sendai, 980-8571,
  Japan}

\maketitle

\begin{abstract}
  A novel interferometric method for absolute beam energy measurement
  is under development at MAMI. At the moment, the method is tested
  and optimized at an energy of 195\,MeV. Despite the very small
  statistical uncertainty of the method, systematic effects have
  limited the overall accuracy.  Recently, a measurement has been
  performed dedicated to the evaluation of these effects. This report
  comprises a description of the method and results of the recent data
  taking period.
\end{abstract}

\section{Experimental setup}
\begin{figure}
  \centering
  \includegraphics[width=\textwidth/2]{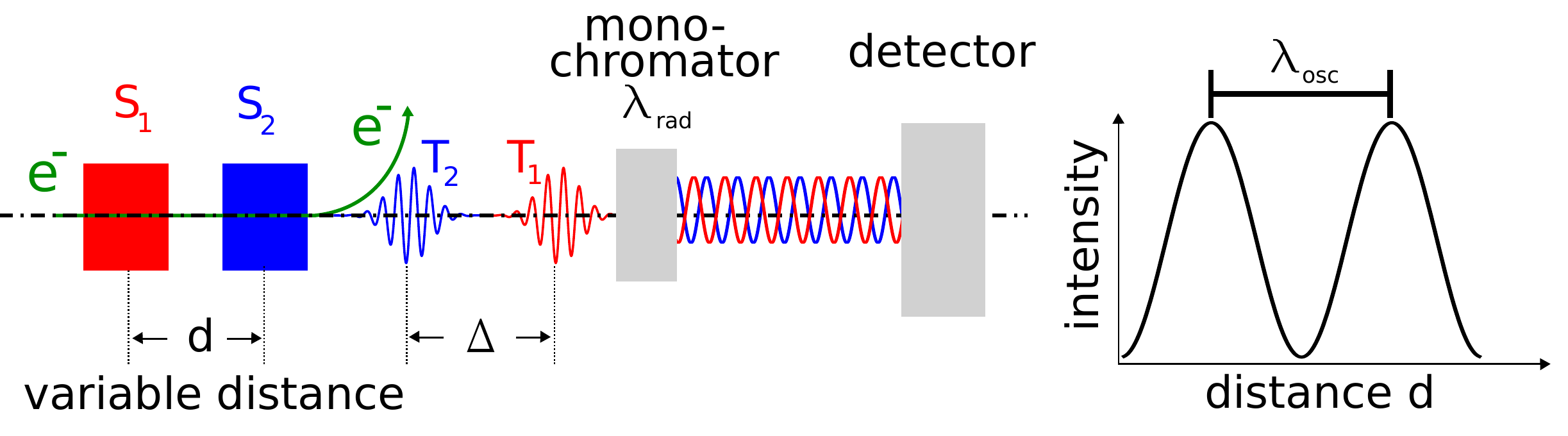}
  \caption{Schematic drawing (not to scale) of the novel method for
    absolute beam energy measurements comprising two spatially
    separated sources of coherent light at an electron beam, an
    optical interferometer system, and an example of the intensity
    oscillation as function of source distance. Relativistic electrons
    ($e^-$) pass through the two sources ($S_1$ and $S_2$ separated
    along the axis by a variable distance $d$) and produce wave trains
    of coherent light ($T_1$ and $T_2$ separated by a difference
    $\Delta$). A monochromator serves as Fourier analyzer of the wave
    trains and a position sensitive optical detector is used to
    observe the interference. The intensity for a selected wavelength
    $\lambda_{\text{rad}}$ will show a periodical variation with $d$,
    its oscillation length $\lambda_{\text{osc}}$ is directly related
    to the beam energy $\gamma = \sqrt{ \lambda_{\text{osc}} /
      2\lambda_{\text{rad}}}$ when observed on-axis.}
  \label{fig:expSetup}
\end{figure}

The method is based on interferometry with two spatial separated light
sources driven by relativistic electrons~\cite{Dambach,Lauth}.  The
basic idea will be explained by means of the schematic experimental
setup shown in Fig.~\ref{fig:expSetup}. An electron beam with Lorentz
factor $\gamma$ passes a pair of undulators S$_1$ and S$_2$ separated
by a distance $d$. The succession of the wave trains T$_1$ and T$_2$
at the exit of the undulator pair is opposite to the order of the two
sources because the electron velocity $v$ is slower than the speed of
light.  These trains are separated along the axis by the distance
\begin{align}
  \Delta(\theta,d) = \left( \frac{2+K^2}{4\gamma^2} +
    \frac{\theta^2}{2} \right)L_U + \left(
    \frac{1}{2\gamma^2}+\frac{\theta^2}{2} \right)d\,,
\end{align}
which is a linear function in $d$ with the linear term only dependent
of the Lorentz factor $\gamma$ and the observation angle $\theta$ with
respect to the electron beam direction.  $L_U \simeq n \lambda_U$ is
the length of the undulator with $\lambda_U$ being the undulator
period, $n$ the number of periods, and $K$ is the undulator
parameter. The undulators act as sources for the emission of coherent
light with the amplitudes ($A$) of the two wave trains having a phase
difference of $\phi(\theta,d) = 2\pi \Delta(\theta,d) /
\lambda_{\text{rad}}$ for a given wavelength of the radiation. The
following equation describes the intensity $I = A^2$ of the two
interfering sources:
\begin{align}
  I(\theta,d) = |A_1|^2 + |A_2|^2 + 2|A_1||A_2| \cos {2\pi
    \Delta(\theta,d)/\lambda_{\text{rad}}}\,.
\end{align}
A monochromator can serve as a Fourier analyzer of the wave trains. If
both wave trains interfere in a position sensitive detector and $d$ is
varied by moving one of the sources, then the revolving phase
$\phi(\theta,d)$ can be observed as intensity oscillations. For a
wavelength $\lambda_{\text{rad}}$ given by the monochromator and
on-axis observation at $\theta = 0$, the oscillation length
$\lambda_{\text{osc}}$ is directly related to the $\gamma$ factor:
\begin{align}
  \lambda_{\text{osc}} = 2\gamma^2
  \frac{\lambda_{\text{rad}}}{1+\gamma^2 \Theta^2}
  \label{eq:OscWithTheta}
\end{align}
Both, $\lambda_{\text{osc}}$ as well as $\lambda_{\text{rad}}$, can be
measured with very high precision.  The method is independent of the
nature of the emission process, provided that the produced light is
coherent. In a pioneering experiment the proof-of-concept of this
method was demonstrated.

\section{Synchrotron radiation measurements at MAMI}

\begin{figure}
  \centering
  \includegraphics[height=.5\textheight]{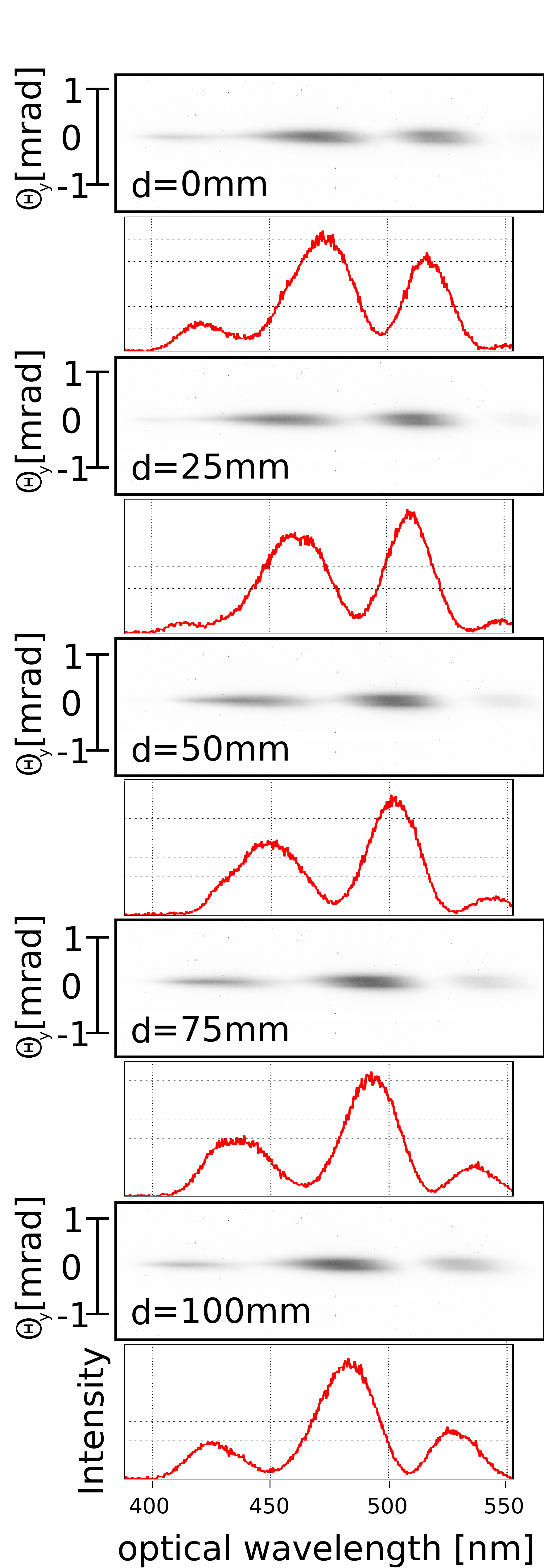}
  \caption{Evolution of the synchrotron spectrum taken with the CCD
    camera for a variation of the undulator distance $d$ from 0 to
    100\,mm. The 16-bit grayscale inverted images show areas of
    600\,px $\times$ 160\,px of the CCD.  The rows (horizontal
    direction) resolve the wavelength, the columns (vertical
    direction) image the non-dispersive angle $\theta_y$ of the
    radiation. The spectra below the images show the intensity
    distributions for the single row at $\theta_y = 0$.}
  \label{fig:Figure2}
\end{figure}

\begin{figure}[tbp]
  \centering
  \includegraphics[height=.3\textheight]{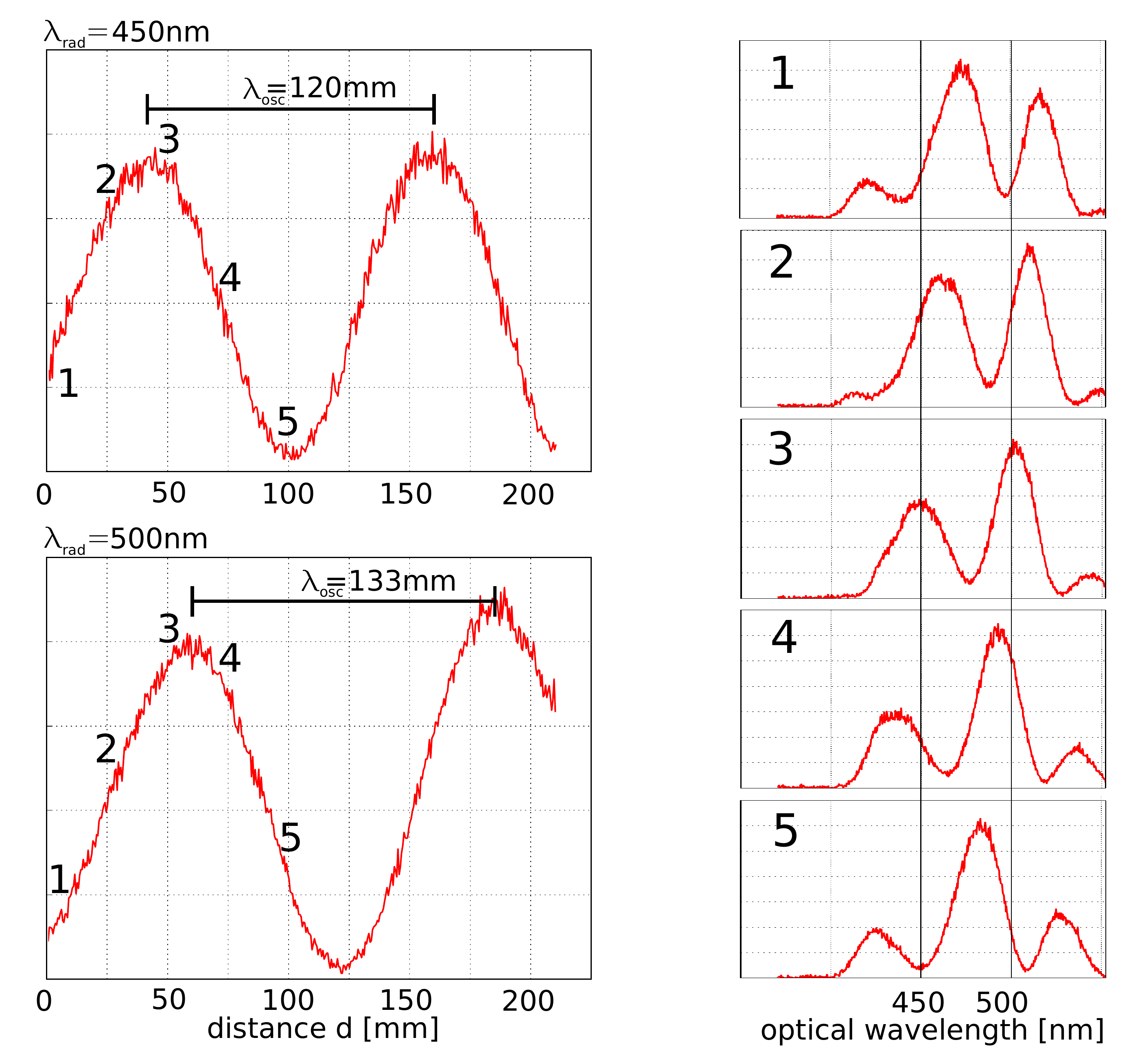}
  \caption{Right: Evolution of the intensity distributions for the
    single-pixel row at $\theta_y = 0$ with the distance $d$. Left:
    Evolution of the intensities for two selected wavelengths at
    450\,nm and 500\,nm. The intensities in each pixel of one CCD row
    show a periodic variation. The numbers describe the correspondence
    between the spectra on the right and the interference oscillations
    on the left.  The oscillation length increased by approximately
    11\,\% when selecting a $500/450\approx 1.11$ larger
    wavelength. The beam energy was approximately 195\,MeV.}
  \label{fig:Figure3}
\end{figure}

The two undulators were used as synchrotron radiation sources at the
MAMI electron beam in two consecutive beam-times to verify their
proposed application for absolute beam energy measurements. The beam
energy was approximately 195\,MeV. Fig.~\ref{fig:Figure2} shows a
series of images taken with the CCD camera for five different
undulator distances in steps of $25\,$mm and corresponding synchrotron
spectra. For the determination of the oscillation length, the distance
was varied in steps of 0.5\,mm, so that approximately 200 separate
spectra could be analyzed. This evolution of the intensity
distribution with distance is shown in Fig.~\ref{fig:Figure3} for
these five positions together with the full intensity variation for
two selected wavelengths near 450\,nm and 500\,nm.  As expected, the
intensity for each wavelength in the spectrum undergoes a periodic
oscillation as the undulator moves.  It was verified that the
oscillation length increases proportional to the wavelength. Since the
statistical noise of the CCD increases with intensity, the oscillation
curves show larger fluctuations near the maximum positions as compared
to the minimum positions. In the second beam-time the measurements
have been continued with a different optical spectrometer utilizing a
grating monochromator that provided an approximate 60 times larger
dispersion.

\begin{figure}
  \centering
  \includegraphics[height=.3\textwidth]{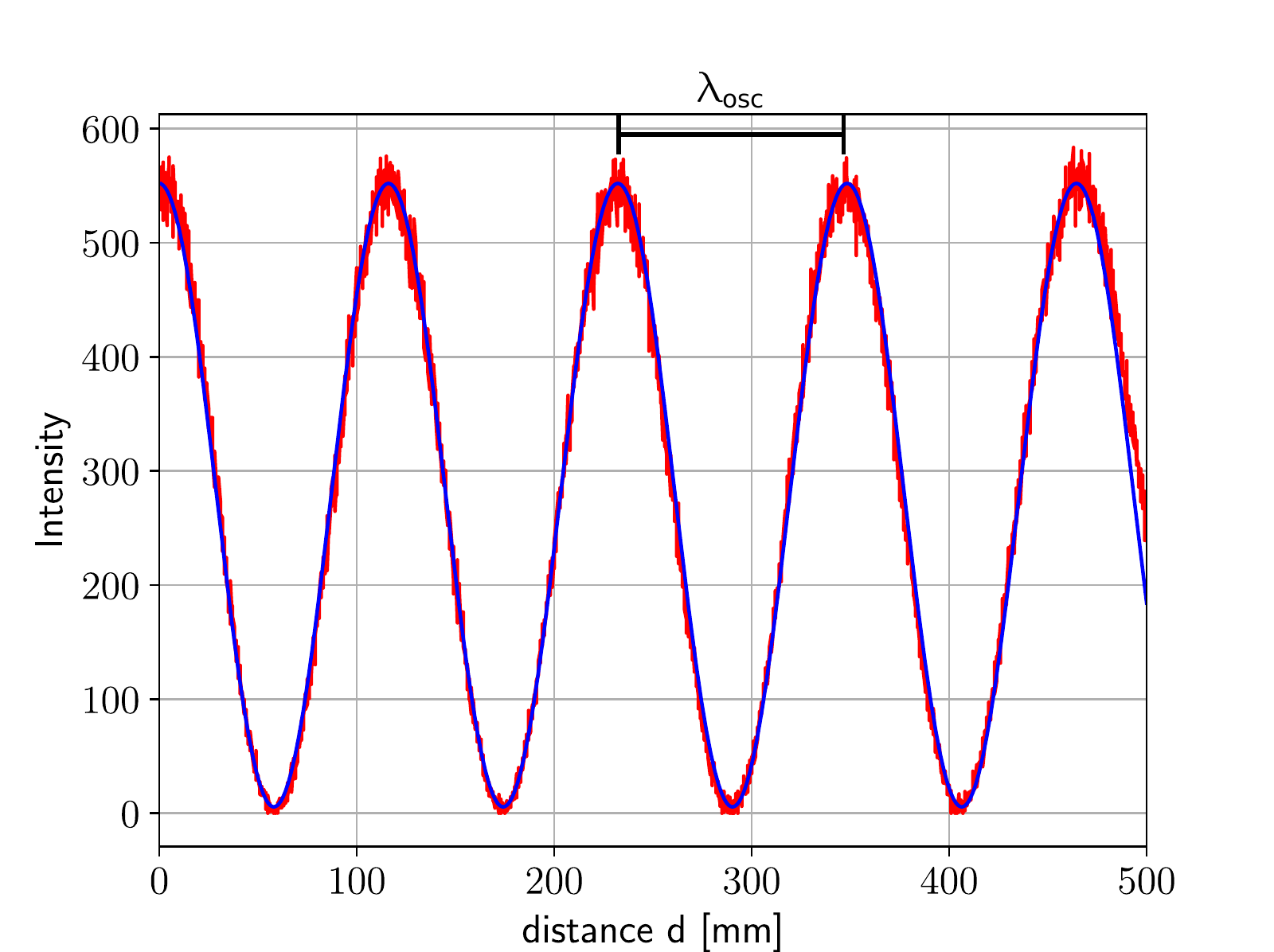}
  \caption{ Intensity as a function of distance d taken from a single
    CCD pixel corresponding to one exemplary wavelength and minimum
    $\Theta_y$ (red data points). The data was fit with a sine
    function (blue line) from which an oscillation length was
    determined.}
\end{figure}

For the determination of the electron beam energy, at first the
single-pixel rows of the CCD images at $\theta_y = 0$ were determined,
which lie in the vertical center of the synchrotron radiation
cone. From their oscillation the interference observable
$\sqrt{\lambda_{\text{osc}}/2\lambda_{\text{rad}}}$ could then be
extracted for each wavelength band corresponding to a single CCD
pixel. This observable is representing the Lorentz factor $\gamma$ of
the electron beam plus contributions depending on differences between
light emission and observation angles. With the accepted wavelength
band of the monochromator covering all 2\,328 CCD pixels in the
horizontal direction of the camera, the same number of simultaneous
determinations of interference oscillations could be performed in one
measurement run.

\begin{figure}
  \centering
  \includegraphics[height=.3\textwidth]{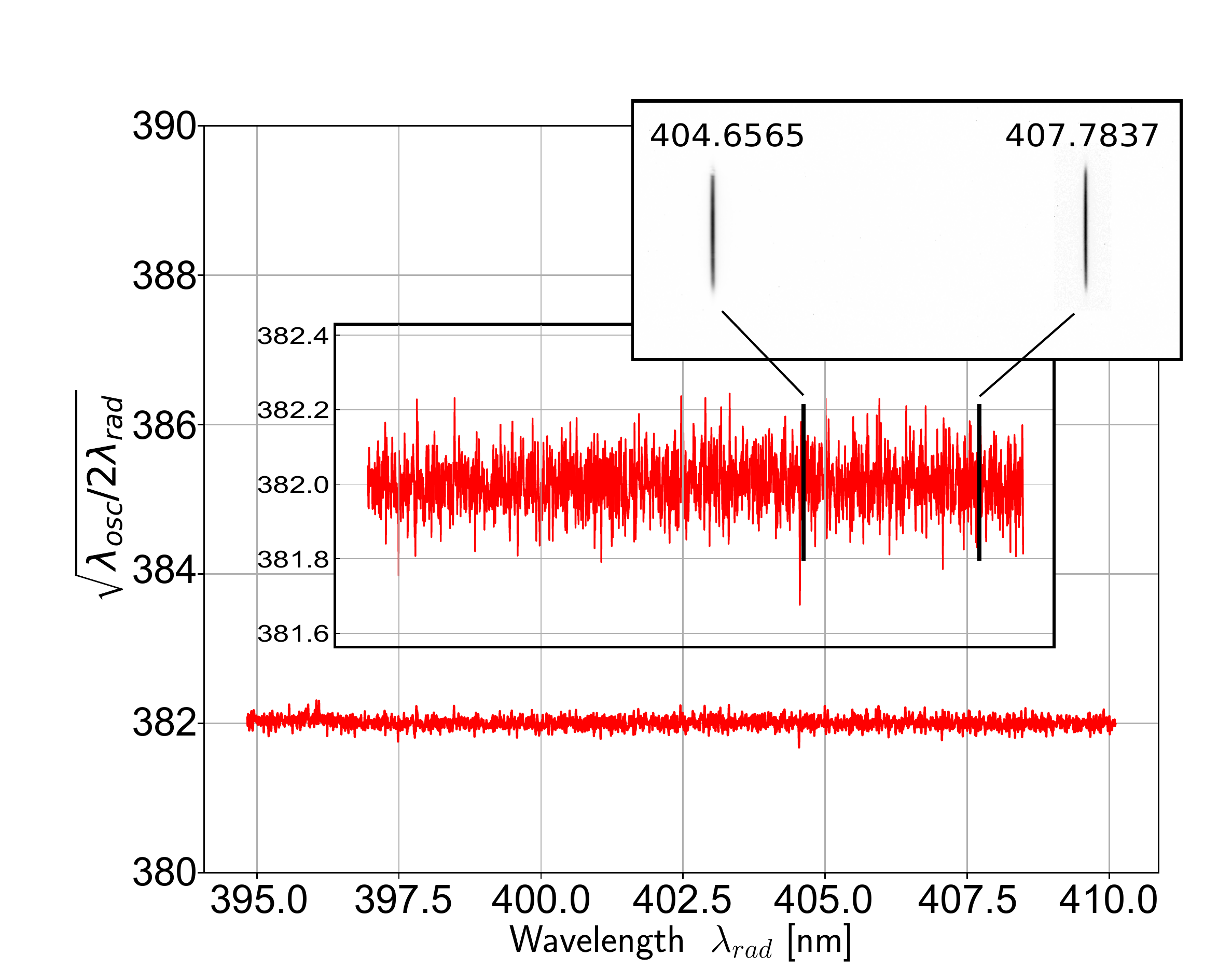}
  \caption{Interference observable
    $\sqrt{\lambda_{\text{osc}}/2\lambda_{\text{rad}}}$ as
    simultaneously observed for 2\,328 different wavelength bands at
    $\theta_y = 0$. The values were deduced from the interference
    oscillations during one measurement run. The insert shows a
    magnification of the scale so that the point-to-point fluctuations
    in the data are visible. No wavelength dependence was
    observed. The interference observable represents the Lorentz
    factor $\gamma$ of the electron beam plus off-axis
    contributions. The optical interferometer system has been
    calibrated with Hg emission lines at 404.6565 and 407.7837\,nm
    whose positions are indicated by the vertical markers.}
\end{figure}

\begin{figure}
  \centering
  \includegraphics[width=0.45\textwidth]{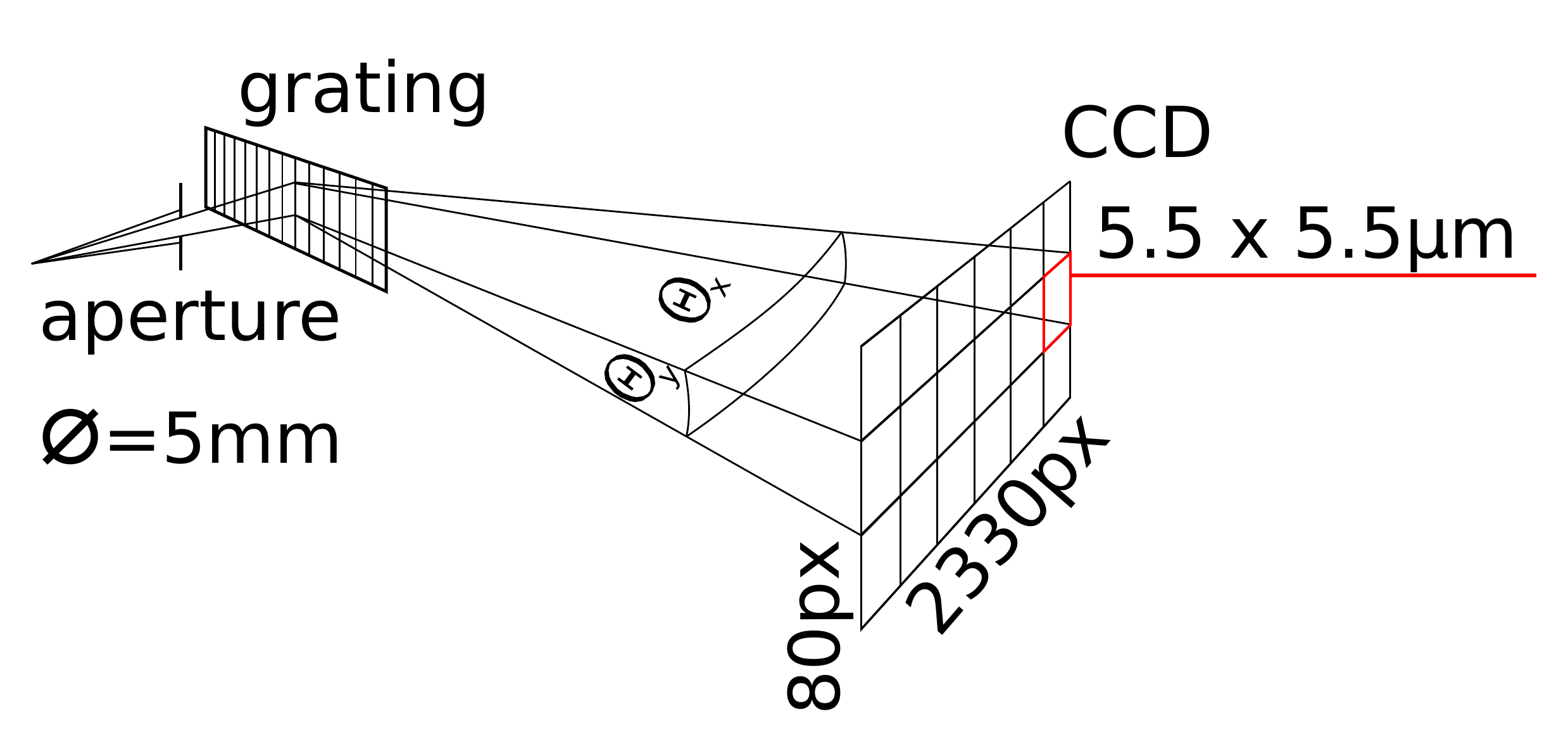}
  \caption{ Schematic drawing of the CCD setup (not to scale). The
    2\,328 rows of the CCD resolved the light spectrum in the
    dispersive direction. An image of the entrance aperture could be
    observed in the non-dispersive direction with 80 of the 1\,750
    columns of the CCD.}
\end{figure}

\section{Sources of systematic uncertainties}
\begin{figure}
  \centering
  \includegraphics[width=0.45\textwidth]{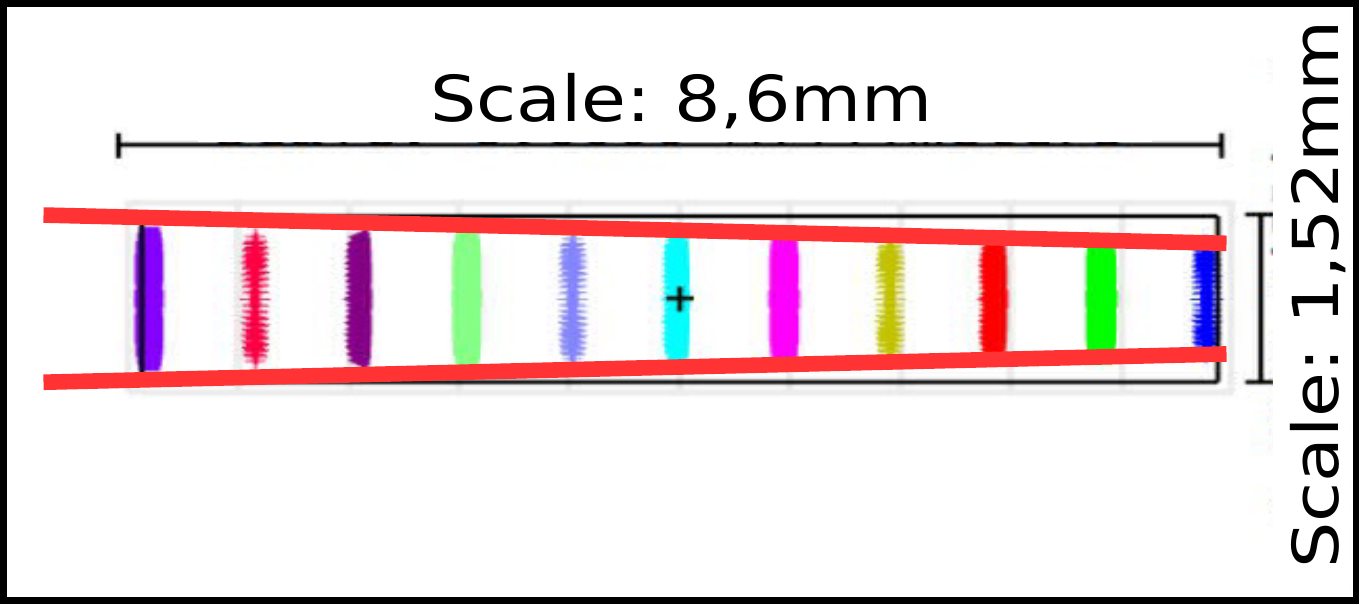}
  \caption{Simulated spectrogram of a Czerny Turner monochromator with
    identical properties. Each bar of false color represents one
    wavelength that has been traced through the monochromator. The
    horizontal dimension of 8.6\,mm corresponds to the extends of the
    CCD-chip. If uniform dispersion were applicable for all pixel
    lines, all wavelengths should be rendered to equal heights. On the
    contrary, a weak tapering is detected. Sketch prepared
    by~\cite{Sablowski}.}
  \label{fig:SablowskiSketch}
\end{figure}

In Eq.~(\ref{eq:OscWithTheta}) the observation angle $\Theta$ gives a
non-vanishing contribution to the interference variable. It has
components in the vertical- and in the horizontal direction,
$\Theta_{x,y}$.  Minimization of both is required in order to obtain
the beam energy correctly. Fig.~\ref{fig:SablowskiSketch} shows an
optical simulation of the Czerny Turner monochromator.  Images of a
series of different wavelengths, each spaced by 1\,nm, are shown in
the spectrum. The simulation demonstrates that each wavelength is
imaged within a different height. This non-uniformity needs to be
considered if off-axis spectra are taken for the in-situ alignment
process.

\section{Conclusion}
The interference of synchrotron radiation from two undulators was
measured over a range of up to 500 mm by analyzing the spectrum with a
monochromator. No coherence loss was observed. In the future, to
achieve high accuracy in the evaluation of the two dimensional
interferogram, the non-uniformity of the Czerny Turner monochromator
has to be compensated.



\end{document}